# Novel Bounded Binary-Addition Tree Algorithm for Binary-State Network Reliability Problems

Wei-Chang Yeh
Integration and Collaboration Laboratory
Department of Industrial Engineering and Engineering Management
National Tsing Hua University
yeh@ieee.org

*Abstract* — Many network applications are based on binary-state networks, where each component has one of two states: success or failure. Efficient algorithms to evaluate binary-state network reliability are continually being developed. Reliability estimates the probability of the success state and is an effective and popular evaluation technique for binary-state networks. Binary-addition tree (BAT) algorithms are frequently used to calculate the efficiency and reliability of binary-state networks. In this study, we propose a novel, bounded BAT algorithm that employs three novel concepts: the first connected vector, the last disconnected vector, and super vectors. These vectors and the calculations of their occurrent probabilities narrow the search space and simplify the probability calculations to reduce the run time of the algorithm. Moreover, we show that replacing each undirected arc with two directed arcs, which is required in traditional direct methods, is unnecessary in the proposed algorithm. We call this novel concept the undirected vectors. The performance of the proposed bounded BAT algorithm was verified experimentally by solving a benchmark set of problems.

**Keywords**: Network; Binary-state; Reliability; Binary Addition Tree Algorithm (BAT); Bound

## 1. INTRODUCTION

With the rapid growth of advanced technology, various networks have emerged and become more diversified and powerful. Such networks include the Internet of Things [1], 4G/5G telecommunications [2], social networks [3], deep learning [4, 5], cloud/fog/edge computing [6], and smart wireless sensor networks [7, 8]. Modern networks and traditional networks (e.g., water, gas, electric, and telephone) are integral to daily life for almost all humans and industries (e.g., manufacturing, business, and supply chain) worldwide [9, 10].

Network reliability is the most well-known technique for estimating the probability that the network is still operational under predefined conditions, including signal quality [11], budget [12], transition time [13], transition rate [14, 15], weight [16], volume [17], production numbers [18], and flow [19].

A binary-state network, where the state of each arc is either functioning or failed, forms the basis of many networks[20, 21, 22, 23], including the multi-state flow networks of which each arc can have more than two states [24, 25, 26, 27, 28, 29, 30, 31]; information networks of which the flow is unsatisfied the conservation law [32, 33]; multi-commodity multi-state flow/information networks which allow different types of flows in multi-state flow/information networks [34, 35]; and multi-distribution multi-commodity multi-state flow/information networks of which each arc can have more than one probability distribution multi-commodity multi-state flow/information networks [36]. Improving the efficiency of algorithms that calculate binary-state network reliability can enhance reliability assessment in diverse networks.

Calculating binary-state network reliability is NP-hard and #P-Hard [20, 21]. Scholars have begun to research the reliability of traditional networks (e.g., transportation and communication) from different aspects, such as approximated and exact reliability. Many approximated reliability methods are based on the Monte Carlo simulation [5, 37] or are bounded [38]. Such methods can be implemented in more extensive networks, for which the rapidity of calculating the reliability estimate is more important than its accuracy [23].

Exact reliability methods can be categorized as direct [20, 21] (e.g., state-space algorithm [44, 45] and the binary-decision diagram (BDD) [39, 40, 41]) and indirect [1, 4, 5, 8, 11, 12, 19, 25, 26, 29, 30, 31, 33, 34, 35, 36, 38, 41, 42, 43]. Indirect methods are based on minimal cuts (MCs) [41, 42, 43] or minimal paths (MPs) [1, 4, 5, 8, 11, 12, 19, 25, 26, 29, 30, 31, 33, 34, 35, 36, 38] and need to solve an NP-hard problem (or a #P-Hard problem) to obtain all the MPs or MCs [20, 21].

Subsequently, another NP-hard problem must be solved by an inclusion-exclusion technique (IET) [24, 27] or a sum of disjoint products method (SDP) [28, 29, 30] using the identified MPs or MCs.

Direct methods are based on connected or disconnected vectors. The graph related to a connected vector has a connection between nodes 1 and $n$. If there is no such connection, the vector is disconnected. The state-space algorithm of direct methods is less efficient than identifying the MPs or MCs of indirect methods, and the BDD is challenging to code [3, 40, 41]. Therefore, indirect methods based on MCs or MPs are more popular than direct methods [1, 4, 5, 8, 11, 12, 19, 25, 26, 29, 30, 31, 33, 34, 35, 36, 38, 41, 42, 43]. However, one study suggests that the BAT direct method [44] solves one NP-hard problem to obtain all connected vectors or disconnected vectors and is more efficient than using indirect methods and solving two NP-hard problems [20, 21].

Big data is increasing the size of modern networks. Consequently, more efficient exact-reliability algorithms are needed for solving larger-scale problems. The original BAT proposed by Yeh is a direct method for calculating binary-state network reliability [44]. The BAT is more efficient, more straightforward, less challenging to code, and more customizable than traditional methods, including the DFS, BFS, and UGFM [44]. Therefore, the BAT has been applied to solving the wildfire problem [46], computer virus problem [47], and the theory of inventive problem solving (TRIZ) problem and has expanded to encompass two- [46] and three-dimensions [47].

The purpose of this study was to propose a new bounded BAT to calculate the binary-state reliability without using MPs and MCs. Bounded BAT introduces novel concepts (undirected vectors, the first connected vector, the last disconnected vector, and super vectors) and corresponding probability calculations that make it faster than a traditional indirect BAT [44].

The remainder of this paper is organized as follows: Section 2 presents all required acronyms, notations, nomenclature, and assumptions used in the proposed bounded BAT. Section 3 reviews the fundamentals of the proposed bounded BAT: a novel forward minimum cut, a novel backward minimum cut, and the BAT [44]. Section 4 introduces the most innovative parts of the proposed bounded BAT, including the novel concepts: the undirected vectors, first connected vector, last disconnected vector, and super vectors. Section 5 provides a simple way to calculate the (occurrent) probabilities of the super vectors and the vectors before the first connected vector or after the last disconnected vector. Section 6 presents the bounded BAT computer pseudo-code, a demonstration, a discussion of its time complexity, and experimental results. Section 7 concludes the paper.

## 2. ACRONYMS, NOTATIONS, NOMENCLATURE, AND ASSUMPTIONS

All required acronyms, notations, nomenclature, and assumptions are provided in this section.

### 2.1 Acronyms

- MP/MC: minimal path/cut
- BAT: binary-addition tree algorithm
- Bounded BAT: proposed new BAT
- BFS: breadth-first-search method
- DFS: depth-first-search method
- UGFM: universal generating function methodology
- IET: inclusion–exclusion technology
- SDP: sum-of-disjoint products method

### 2.2 Notations

- $/\bullet/$: number of elements in set $\bullet$
- $\Pr(\bullet)$: success probability of event $\bullet$
- $n$: number of nodes
- $m$: number of arcs
- $V$: set of nodes $V = \{1, 2, …, n\}$
- $E$: set of arcs $E = \{a_1, a_2, …, a_m\}$

$a_k$: $k^{th}$ arc in $E$

$\varepsilon_{i,j}$: directed arc between nodes $i$ and $j$

$e_{i,j}$: undirected arc from nodes $i$ to $j$

$\mathbf{D_b}$: An arc binary-state distribution function that lists the probability of a functioning state for each arc. For example, Table 1 is an arc binary-state distribution function of seven arcs.

**Table 1.** Example of arc state distribution function.

| $i$ | $\Pr(a_i)$ | $i$ | $\Pr(a_i)$ |
|---|---|---|---|
| 1 | 0.98 | 5 | 0.75 |
| 2 | 0.80 | 6 | 0.90 |
| 3 | 0.85 | 7 | 0.88 |
| 4 | 0.95 | | |

$G(V, E)$: A graph with sets of nodes $V$, arcs $E$, source node 1, and a sink node $n$. For example, Fig. 1 shows a graph for source node 1, sink node 4, $V = \{1, 2, 3, 4, 5\}$, and $E = \{a_i \mid i = 1, 2, \ldots, 7\}$.

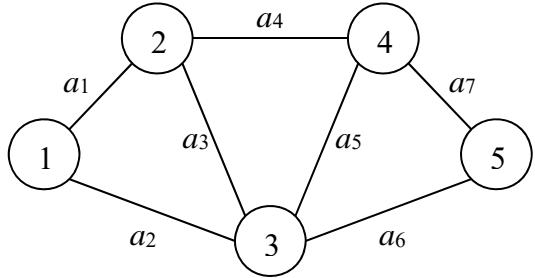

**Figure 1.** Example network

$G(V, E, \mathbf{D_b})$: A binary-state network with graph $G(V, E)$ and arc binary-state distribution function $\mathbf{D_b}$. The graph in Fig. 1 is a binary-state network $G(V, E, \mathbf{D_b})$ after obtaining $\mathbf{D_b}$, where $\mathbf{D_b}$ is presented in Table 1.

$R$: reliability of $G(V, E, \mathbf{D_b})$

$X$: (state) vector

$X_{FC}$: first connected (state) vector in the proposed bound BAT

$X_{LD}$: last disconnected (state) vector in the proposed bound BAT

$X(i)$: state of the $i^{th}$ coordinate in $X$

$X(a_i)$: state of $a_i$ in $X$

$X(1:i)$: A subvector that includes the first $i$ coordinates with the same order and values of $X$

$\Pr(X(a_i))$: An occurrence probability of $a_i$ in $\mathbf{D_b}$ when its state is $X(a_i)$

$\Pr(X)$: $\Pr(X) = \Pr(X(a_1)) \times \Pr(X(a_1)) \times \ldots \times \Pr(X(a_m))$

$G(X)$: Subgraph $G(X) = G(V, \{a \in E \mid \text{for all } a \text{ with } X(a) = 1\})$, corresponding to state vector $X$ in $G(V, E)$. $G(X)$ is illustrated in Fig. 2, where $X = (0, 1, 1, 1, 0, 1, 1)$.

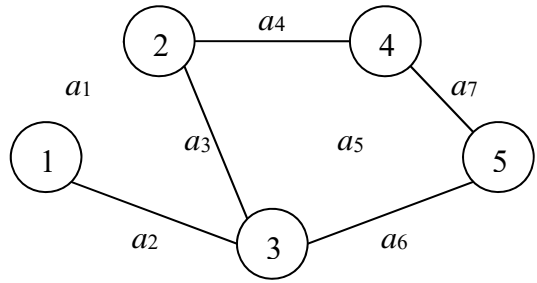

**Figure 2.** $G(X)$ and $X = (0, 1, 1, 1, 0, 1, 1)$ in Fig. 1.

$n_p$: The number of arcs in the shortest paths from nodes 1 to $n$. For example, $n_p = 2$ in Fig. 1 because $\{a_1, a_4\}$ and $\{a_2, a_5\}$ are the shortest paths.

$n_c$: The number of arcs in any minimum cut between node 1 and node $n$. For example, $n_c = 2$ in Fig. 1 because $\{a_1, a_2\}$ and $\{a_4, a_5\}$ are the minimum cuts.

$A \leq B$: $A(a_i) \leq B(a_i)$ for all $i = 1, 2, \ldots, m$, e.g., $(1, 0, 0, 0, 1) \leq (1, 0, 1, 0, 1)$

$A < B$: $A \leq B$ for all $i = 1, 2, \ldots, m$, and $A(a_j) < B(a_j)$ for at least one $j = 1, 2, \ldots, m$. For example, $(1, 2, 3, 4, 5) < (1, 2, 3, 4, 6)$

$A \ll B$: $A \ll B$ if vector $A$ is obtained after $B$ in the BAT. For example, $(0, 1, 0, 0, 1) \ll (1, 0, 1, 0, 0)$. Note that $(0, 1, 0, 0, 1)$ is either less than or greater than $(1, 0, 1, 0, 0)$. That is, both $(0, 1, 0, 0, 1) < (1, 0, 1, 0, 0)$ and $(1, 0, 1, 0, 0) < (0, 1, 0, 0, 1)$ are not held.

$A \ll B$: $A \ll B$ or $A = B$.

## 2.3 Nomenclature

| | |
|---|---|
| Reliability: | The probability that node 1 is connected in succession to node $n$. |
| Shortest Path: | An arc subset and a path from nodes 1 to $n$ such that the total number of its arcs is minimized. |
| Minimum Cut: | An arc subset and a cut between nodes 1 and $n$ such that the total number of its arcs is minimized. |
| Directed vector: | A state vector $X$ presents the state of arcs, each of which must be replaced by two directed arcs in opposite directions; that is, an undirected arc $\varepsilon_{i,j}$ is replaced with directed arcs $e_{i,j}$ and $e_{j,i}$. |
| Undirected vector: | A state vector $X$ presents the state of arcs without changing the undirected arcs into two directed arcs. |
| Connected vector: | A state vector $X$ is connected if nodes 1 and $n$ are connected in $G(X)$. |
| Disconnected vector: | A state vector $X$ is disconnected if nodes 1 and $n$ are disconnected in $G(X)$. |
| First connected vector: | The connected vector $X^*$ such that $X$ is a disconnected vector for all state vectors $X \ll X^*$. |
| Last disconnected vector: | The disconnected vector $X^*$ such that $X$ is a connected vector for all state vectors $X^* \ll X$. |

## 2.4 Assumptions

1. The network is connected without any parallel arcs or loops.

2. All nodes are perfectly reliable, and each arc is either functioning or has failed.

3. The probability of an arc is statistically independent and has a predefined distribution.

## 3. FORWARD MINIMUM CUTS, BACKWARD MINIMUM CUTS, AND THE BAT

The proposed bounded BAT is a conventional BAT [44] combined with novel forward and backward minimum cuts to derive novel vectors, including the first connected vector, the last disconnected vector, and super vectors, and calculate the reliability of binary state networks more efficiently. Therefore, the proposed forward minimum cut, the proposed backward minimum cut, and the traditional BAT are discussed in this section.

### 3.1 Forward Minimum Cuts and Backward Minimum Cuts

Cuts and paths are arc subsets. Removing cuts ensures that no path exists from node 1 to $n$ (i.e., nodes 1 and $n$ are disconnected). A shortest path $p$ and a minimum cut $c$ are a path and a cut between nodes 1 and $n$ such that the total number of arcs is minimum, respectively. For example, there are two minimum cuts $\{a_1, a_2\}$ and $\{a_6, a_7\}$ and one shortest path $\{a_2, a_6\}$ in Fig. 1.

The Dijkstra algorithm is one of the most popular for finding the shortest path problem from node 1 to $n$ with time complexity $O(n^2)$ [48]. Among the minimum cut-related algorithms, the Stoer-Wagner deterministic minimum cut algorithm is the most efficient and can find the minimum cut in $O(mn + n^2 \log n)$ [49].

Forward minimum cuts and backward minimum cuts are two novel concepts proposed in this study. These minimum cuts are simple arc subsets and are formatted into sub-vector forms. Subvector $X(1:k)$ is a forward minimum cut if $G(X(1:k))$ is a disconnected graph and $G(X(1:k-1))$ is not disconnected. Subvector $X(k:m)$ is a backward minimum cut if $G(X(k:m))$ is a disconnected graph and $G(X(k-1:m))$ is not disconnected.

### 3.2 Connected and Disconnected Vectors

A vector $X$ is called a connected vector if there is a path from nodes 1 to $n$ in its related network $G(X)$. If there is no such path, it is called a disconnected vector. Therefore, a vector is connected or disconnected. For example, (0, 0, 1, 1, 1, 1, 1) and (1, 1, 1, 1, 1, 0, 0) are the disconnected vectors related to the minimum cuts $\{a_1, a_2\}$ and $\{a_6, a_7\}$, respectively, and (0, 1, 0, 0, 0, 1, 0) is the connected vector of the shortest path $\{a_2, a_6\}$.

### 3.3 BAT

The BAT proposed by Yeh [44] uses a new heuristic search method and is more efficient than DFS. This BAT is also more effective than BFS [44] and UGFM [50] because the latter methods generate incorrect results because of computer memory overflow. Furthermore, BAT is simple to understand, easy to code, and customizable [44, 45, 46, 47, 50].

In a binary vector, the value of each coordinate is either 0 or 1. Assume that a search process updates a binary vector using a procedure that iteratively moves through each coordinate from last to first. At the beginning of the updating process, the binary vector is called "vector zero."

1. If the current coordinate is 0, *it is changed to 1 and a new binary vector is generated. For example, $X$* = (1, 0, 0) is updated to $X^*$ = (1, 0, 1).

2. If the current coordinate is 1, then it *is changed to 0. Go to the next coordinate to repeat the above two steps. For example, $X^*$* = (1, 0, 1) is changed to $X^\#$ = (1, 1, 0).

The original BAT pseudo-code is delivered as follows [44]:

**STEP A0.** Replace each undirected arc with two directed arcs, except for the arcs with an endpoint at node 0 or a node adjacent to node $n$. Let $m^*$ be the number of arcs after the transformation; let SUM = 0; let $X$ be a vector zero, and let $i = m^*$.

**STEP A1.** If $X(a_i) = 1$, let SUM = SUM − 1, $X(a_i) = 0$, and go to STEP A3.

**STEP A2.** Let SUM = SUM + 1 and $X(a_i) = 1$.

**STEP A3.** Let $i = i − 1$ and return to STEP A1 if $i > 1$.

**STEP A4.** If SUM = $m^*$, halt; otherwise, let $i = m^*$ and return to STEP A1.

In STEP A0, vector zero is the original vector used to generate all the vectors in the BAT. Note that the reason we need to replace the undirected arcs with two directed arcs is explained in Section 4.1. STEPs A1 to A4 form a process that generates a new vector. STEPs A1 to A4 are repeated for each vector, and the last generated vector is vector one. The value of the current coordinate (i.e., the $i$th coordinate) is changed from 0 to 1 or 1 to 0, *as shown in STEPs A1 and A3, respectively,* to reduce the computer burden time. *Subsequently,* if the $i$th coordinate is changed from 1 to 0, *we need to go to the $(i-1)^{th}$ coordinate and STEP A1, and repeat the same procedure until the current* coordinate is changed from 0 to 1.

The stopping criteria is that all coordinates are one (i.e., SUM = $m^*$), and STEP A4 tests whether this condition has been met. Note that each new $X$ generated in STEP A1 can be used to determine corresponding values (e.g., probability, cost, time, or any predefined function). In addition, there is no need to store the vectors that *are generated before the current X is updated*.

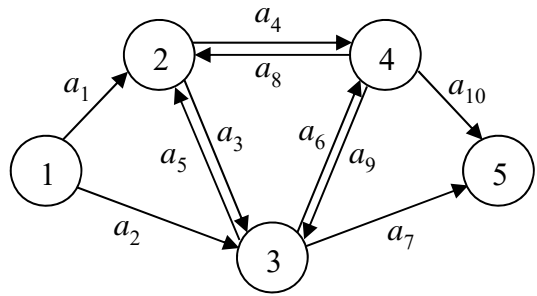

**Figure 3**. Graph from Fig. 1 in which undirected arcs are replaced with directed arcs.

For instance, the graph in Fig. 3 is the result of replacing undirected arcs with directed arcs in Fig 1. Let $a_1 = \varepsilon_{1,2}$, $a_2 = \varepsilon_{1,3}$, $a_3 = \varepsilon_{2,3}$, $a_4 = \varepsilon_{2,4}$, $a_5 = \varepsilon_{3,2}$, $a_6 = \varepsilon_{3,4}$, $a_7 = \varepsilon_{3,5}$, $a_8 = \varepsilon_{4,2}$, $a_9 = \varepsilon_{4,3}$, and $a_{10} = \varepsilon_{4,5}$ in Fig. 3. The value of $m^*$ is 10; $X = (0, 0, 0, 0, 0, 0, 0, 0, 0, 0)$, and SUM = 0 in STEP A0. There are $2^{10} = 1024$ different vectors after generating the BAT pseudo-code, and the first 48 vectors are listed in Table 2. Note that each of these 48 vectors is disconnected. Among *the 1024 vectors, 473 vectors are connected, and the rest are disconnected*. $X_{265} = (0, 1, 0, 0, 0, 0, 1, 0, 0, 0)$ is the first connected vector, and $X_{1015} = (1, 1, 1, 1, 1, 1, 0, 1, 1, 0)$ is the last disconnected vector. Moreover, each vector is unique (i.e., there are no duplications).

**Table 2.** Initial 48 vectors obtained from the traditional BAT.

| i | $X_i$ | i | $X_i$ | i | $X_i$ |
|---|---|---|---|---|---|
| 1  | (0, 0, 0, 0, 0, 0, 0, 0, 0, 0) | 17 | (0, 0, 0, 0, 1, 0, 0, 0, 0, 0) | 33 | (0, 0, 0, 0, 0, 0, 0, 0, 0, 0) |
| 2  | (0, 0, 0, 0, 0, 0, 0, 0, 0, 1) | 18 | (0, 0, 0, 0, 1, 0, 0, 0, 0, 1) | 34 | (0, 0, 0, 0, 0, 0, 0, 0, 0, 1) |
| 3  | (0, 0, 0, 0, 0, 0, 0, 0, 1, 0) | 19 | (0, 0, 0, 0, 1, 0, 0, 0, 1, 0) | 35 | (0, 0, 0, 0, 0, 0, 0, 0, 1, 0) |
| 4  | (0, 0, 0, 0, 0, 0, 0, 0, 1, 1) | 20 | (0, 0, 0, 0, 1, 0, 0, 0, 1, 1) | 36 | (0, 0, 0, 0, 0, 0, 0, 0, 1, 1) |
| 5  | (0, 0, 0, 0, 0, 0, 0, 1, 0, 0) | 21 | (0, 0, 0, 0, 1, 0, 0, 1, 0, 0) | 37 | (0, 0, 0, 0, 0, 0, 0, 1, 0, 0) |
| 6  | (0, 0, 0, 0, 0, 0, 0, 1, 0, 1) | 22 | (0, 0, 0, 0, 1, 0, 0, 1, 0, 1) | 38 | (0, 0, 0, 0, 0, 0, 0, 1, 0, 1) |
| 7  | (0, 0, 0, 0, 0, 0, 0, 1, 1, 0) | 23 | (0, 0, 0, 0, 1, 0, 0, 1, 1, 0) | 39 | (0, 0, 0, 0, 0, 0, 0, 1, 1, 0) |
| 8  | (0, 0, 0, 0, 0, 0, 0, 1, 1, 1) | 24 | (0, 0, 0, 0, 1, 0, 0, 1, 1, 1) | 40 | (0, 0, 0, 0, 0, 0, 0, 1, 1, 1) |
| 9  | (0, 0, 0, 0, 0, 0, 1, 0, 0, 0) | 25 | (0, 0, 0, 0, 1, 0, 1, 0, 0, 0) | 41 | (0, 0, 0, 0, 0, 0, 1, 0, 0, 0) |
| 10 | (0, 0, 0, 0, 0, 0, 1, 0, 0, 1) | 26 | (0, 0, 0, 0, 1, 0, 1, 0, 0, 1) | 42 | (0, 0, 0, 0, 0, 0, 1, 0, 0, 1) |
| 11 | (0, 0, 0, 0, 0, 0, 1, 0, 1, 0) | 27 | (0, 0, 0, 0, 1, 0, 1, 0, 1, 0) | 43 | (0, 0, 0, 0, 0, 0, 1, 0, 1, 0) |
| 12 | (0, 0, 0, 0, 0, 0, 1, 0, 1, 1) | 28 | (0, 0, 0, 0, 1, 0, 1, 0, 1, 1) | 44 | (0, 0, 0, 0, 0, 0, 1, 0, 1, 1) |
| 13 | (0, 0, 0, 0, 0, 0, 1, 1, 0, 0) | 29 | (0, 0, 0, 0, 1, 0, 1, 1, 0, 0) | 45 | (0, 0, 0, 0, 0, 0, 1, 1, 0, 0) |
| 14 | (0, 0, 0, 0, 0, 0, 1, 1, 0, 1) | 30 | (0, 0, 0, 0, 1, 0, 1, 1, 0, 1) | 46 | (0, 0, 0, 0, 0, 0, 1, 1, 0, 1) |
| 15 | (0, 0, 0, 0, 0, 0, 1, 1, 1, 0) | 31 | (0, 0, 0, 0, 1, 0, 1, 1, 1, 0) | 47 | (0, 0, 0, 0, 0, 0, 1, 1, 1, 0) |
| 16 | (0, 0, 0, 0, 0, 0, 1, 1, 1, 1) | 32 | (0, 0, 0, 0, 1, 0, 1, 1, 1, 1) | 48 | (0, 0, 0, 0, 0, 0, 1, 1, 1, 1) |

## 4. PROPOSED NOVEL VECTORS

In this section, novel vectors are proposed to eliminate unwanted vectors from the reliability calculation: undirected vectors, the first connected vector, the last disconnected vector, and super vectors.

### 4.1 Undirected Vectors

Traditional direct methods for calculating the reliability of binary-state networks need to use directed vectors (i.e., each undirected arc is replaced with two directed arcs in opposite directions). For example, applying a traditional direct algorithm to the example network in Fig. 1 would require that the graph be changed to the one shown in Fig. 4. Note that the directed arcs $e_{i,1} \in E$ and $e_{n,j} \in E$ can be removed [12, 26]. In other words, applying the method proposed by Yeh means that only the undirected arcs adjacent to node 1 and node $n$ need to be replaced with directed arcs, which reduces the number of state vectors. For example, Fig. 1 can be changed to the graph shown in Fig. 3. Note Fig. 3 does not include the directed arcs $e_{2,1}$, $e_{3,1}$, $e_{5,4}$, and $e_{5,3}$ in the graph shown in Fig. 4.

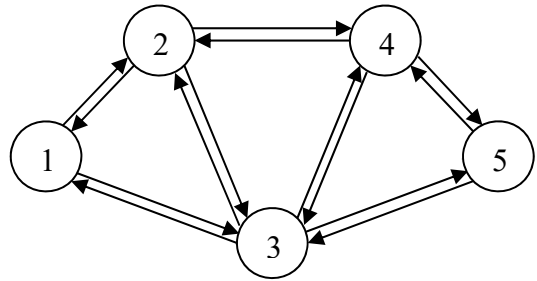

**Figure 4.** Example network to explain the replacement of undirected arcs with directed arcs.

The time complexity of a traditional directed algorithm used to calculate the reliability of a binary-state network with directed vectors is $2^{2|E|} = 2^{2m}$. However, if there is a directed path from nodes 1 to $n$, the state of arcs that are not in the directed path can be ignored.

For example, $\{a_2, a_4, a_5\}$ is a path from nodes 1 to $n$, as shown in Fig. 5, which is identical to Fig. 1 but has different arc labels. The states of $a_1$, $a_3$, $a_6$, and $a_7$ can be ignored, as shown in Fig. 5. Therefore, these undirected arcs do not need to be replaced with two directed arcs in opposite directions. The vectors of undirected arcs that are not replaced are called undirected vectors.

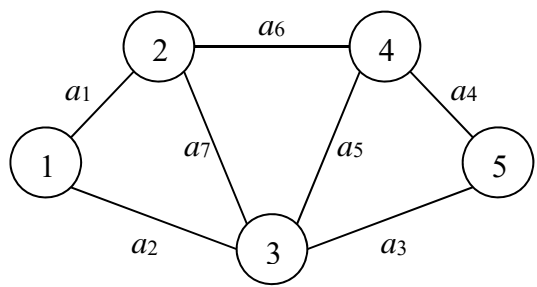

**Figure 5.** Same graph as that shown in Fig. 1 but with different arc labels.

The results from a traditional BAT that uses the proposed undirected vector concept is shown in Table 3. The number of total vectors is decreased from $2^{10} = 1024$ to 128 vectors (i.e., 1/8 of the original number). Therefore, the undirected vector concept is useful for reducing the number of vectors.

**Table 3.** Results of a traditional BAT that uses the proposed undirected vector concept.

| i | $X_i$ | C | $R_i$ | i | $X_i$ | C | $R_i$ | i | $X_i$ | C | $R_i$ | i | $X_i$ | C | $R_i$ |
|---|---|---|---|---|---|---|---|---|---|---|---|---|---|---|---|
| 1 | (0,0,0,0,0,0) | | 128 | 33 | (0,1,0,0,0,0) | | 512 | 65 | (1,0,0,0,0,0) | | 512 | 97 | (1,1,0,0,0,0) | | 2048 |
| 2 | (0,0,0,0,0,1) | | 512 | 34 | (0,1,0,0,0,1) | | 2048 | 66 | (1,0,0,0,0,1) | | 2048 | 98 | (1,1,0,0,0,1) | | 8192 |
| 3 | (0,0,0,0,1,0) | | 512 | 35 | (0,1,0,0,1,0) | | 2048 | 67 | (1,0,0,0,1,0) | | 2048 | 99 | (1,1,0,0,1,0) | | 8192 |
| 4 | (0,0,0,0,1,1) | | 2048 | 36 | (0,1,0,0,1,1) | | 8192 | 68 | (1,0,0,0,1,1) | | 8192 | 100 | (1,1,0,0,1,1) | | 32768 |
| 5 | (0,0,0,1,0,0) | | 512 | 37 | (0,1,0,0,1,0,0) | | 2048 | 69 | (1,0,0,1,0,0) | | 2048 | 101 | (1,1,0,0,1,0,0) | | 8192 |
| 6 | (0,0,0,1,0,1) | | 2048 | 38 | (0,1,0,0,1,0,1) | | 8192 | 70 | (1,0,0,1,0,1) | | 8192 | 102 | (1,1,0,0,1,0,1) | | 32768 |
| 7 | (0,0,0,1,1,0) | | 2048 | 39 | (0,1,0,0,1,1,0) | | 8192 | 71 | (1,0,0,1,1,0) | | 8192 | 103 | (1,1,0,0,1,1,0) | | 32768 |
| 8 | (0,0,0,1,1,1) | | 8192 | 40 | (0,1,0,0,1,1,1) | | 32768 | 72 | (1,0,0,1,1,1) | | 32768 | 104 | (1,1,0,0,1,1,1) | | 131072 |
| 9 | (0,0,0,1,0,0,0) | | 512 | 41 | (0,1,0,1,0,0,0) | | 2048 | 73 | (1,0,0,1,0,0,0) | | 2048 | 105 | (1,1,0,1,0,0,0) | | 8192 |
| 10 | (0,0,0,1,0,0,1) | | 2048 | 42 | (0,1,0,1,0,0,1) | | 8192 | 74 | (1,0,0,1,0,0,1) | | 8192 | 106 | (1,1,0,1,0,0,1) | | 32768 |
| 11 | (0,0,0,1,0,1,0) | | 2048 | 43 | (0,1,0,1,0,1,0) | | 8192 | 75 | (1,0,0,1,0,1,0) | Y | 8192 | 107 | (1,1,0,1,0,1,0) | Y | 32768 |
| 12 | (0,0,0,1,0,1,1) | | 8192 | 44 | (0,1,0,1,0,1,1) | Y | 32768 | 76 | (1,0,0,1,0,1,1) | Y | 32768 | 108 | (1,1,0,1,0,1,1) | Y | 131072 |
| 13 | (0,0,0,1,1,0,0) | | 2048 | 45 | (0,1,0,1,1,0,0) | Y | 8192 | 77 | (1,0,0,1,1,0,0) | | 8192 | 109 | (1,1,0,1,1,0,0) | Y | 32768 |
| 14 | (0,0,0,1,1,0,1) | | 8192 | 46 | (0,1,0,1,1,0,1) | Y | 32768 | 78 | (1,0,0,1,1,0,1) | Y | 32768 | 110 | (1,1,0,1,1,0,1) | Y | 131072 |
| 15 | (0,0,0,1,1,1,0) | | 8192 | 47 | (0,1,0,1,1,1,0) | Y | 32768 | 79 | (1,0,0,1,1,1,0) | Y | 32768 | 111 | (1,1,0,1,1,1,0) | Y | 131072 |
| 16 | (0,0,0,1,1,1,1) | | 32768 | 48 | (0,1,0,1,1,1,1) | Y | 131072 | 80 | (1,0,0,1,1,1,1) | Y | 131072 | 112 | (1,1,0,1,1,1,1) | Y | 524288 |
| 17 | (0,0,1,0,0,0,0) | | 512 | 49 | (0,1,1,0,0,0,0) | Y | 2048 | 81 | (1,0,1,0,0,0,0) | | 2048 | 113 | (1,1,1,0,0,0,0) | Y | 8192 |
| 18 | (0,0,1,0,0,0,1) | | 2048 | 50 | (0,1,1,0,0,0,1) | Y | 8192 | 82 | (1,0,1,0,0,0,1) | Y | 8192 | 114 | (1,1,1,0,0,0,1) | Y | 32768 |
| 19 | (0,0,1,0,0,1,0) | | 2048 | 51 | (0,1,1,0,0,1,0) | Y | 8192 | 83 | (1,0,1,0,0,1,0) | | 8192 | 115 | (1,1,1,0,0,1,0) | Y | 32768 |
| 20 | (0,0,1,0,0,1,1) | | 8192 | 52 | (0,1,1,0,0,1,1) | Y | 32768 | 84 | (1,0,1,0,0,1,1) | Y | 32768 | 116 | (1,1,1,0,0,1,1) | Y | 131072 |
| 21 | (0,0,1,0,1,0,0) | | 2048 | 53 | (0,1,1,0,1,0,0) | Y | 8192 | 85 | (1,0,1,0,1,0,0) | | 8192 | 117 | (1,1,1,0,1,0,0) | Y | 32768 |
| 22 | (0,0,1,0,1,0,1) | | 8192 | 54 | (0,1,1,0,1,0,1) | Y | 32768 | 86 | (1,0,1,0,1,0,1) | Y | 32768 | 118 | (1,1,1,0,1,0,1) | Y | 131072 |
| 23 | (0,0,1,0,1,1,0) | | 8192 | 55 | (0,1,1,0,1,1,0) | Y | 32768 | 87 | (1,0,1,0,1,1,0) | Y | 32768 | 119 | (1,1,1,0,1,1,0) | Y | 131072 |
| 24 | (0,0,1,0,1,1,1) | | 32768 | 56 | (0,1,1,0,1,1,1) | Y | 131072 | 88 | (1,0,1,0,1,1,1) | Y | 131072 | 120 | (1,1,1,0,1,1,1) | Y | 524288 |
| 25 | (0,0,1,1,0,0,0) | | 2048 | 57 | (0,1,1,1,0,0,0) | Y | 8192 | 89 | (1,0,1,1,0,0,0) | | 8192 | 121 | (1,1,1,1,0,0,0) | Y | 32768 |
| 26 | (0,0,1,1,0,0,1) | | 8192 | 58 | (0,1,1,1,0,0,1) | Y | 32768 | 90 | (1,0,1,1,0,0,1) | Y | 32768 | 122 | (1,1,1,1,0,0,1) | Y | 131072 |
| 27 | (0,0,1,1,0,1,0) | | 8192 | 59 | (0,1,1,1,0,1,0) | Y | 32768 | 91 | (1,0,1,1,0,1,0) | Y | 32768 | 123 | (1,1,1,1,0,1,0) | Y | 131072 |
| 28 | (0,0,1,1,0,1,1) | | 32768 | 60 | (0,1,1,1,0,1,1) | Y | 131072 | 92 | (1,0,1,1,0,1,1) | Y | 131072 | 124 | (1,1,1,1,0,1,1) | Y | 524288 |
| 29 | (0,0,1,1,1,0,0) | | 8192 | 61 | (0,1,1,1,1,0,0) | Y | 32768 | 93 | (1,0,1,1,1,0,0) | | 32768 | 125 | (1,1,1,1,1,0,0) | Y | 131072 |
| 30 | (0,0,1,1,1,0,1) | | 32768 | 62 | (0,1,1,1,1,0,1) | Y | 131072 | 94 | (1,0,1,1,1,0,1) | Y | 131072 | 126 | (1,1,1,1,1,0,1) | Y | 524288 |
| 31 | (0,0,1,1,1,1,0) | | 32768 | 63 | (0,1,1,1,1,1,0) | Y | 131072 | 95 | (1,0,1,1,1,1,0) | Y | 131072 | 127 | (1,1,1,1,1,1,0) | Y | 524288 |
| 32 | (0,0,1,1,1,1,1) | | 131072 | 64 | (0,1,1,1,1,1,1) | Y | 524288 | 96 | (1,0,1,1,1,1,1) | Y | 524288 | 128 | (1,1,1,1,1,1,1) | Y | 2097152 |

C: marked "Y" when the vector is connected; if unmarked, it is disconnected.

$R_i$: $R_i = \Pr(X_i) \times 10^7$

Moreover, only the probabilities of the real arcs in the path are needed to calculate the related probability. For example, the probability of the subnetworks included in $\{a_2, a_5, a_7\} = \{e_{1,3}, e_{3,4}, e_{4,5}\}$ is $\Pr(\{a_2, a_5, a_7\}) = \Pr(\{a_2\}) \times \Pr(\{a_5\}) \times \Pr(\{a_7\})$.

The time complexity of any conventional direct method can be reduced from $2^{2m}$ to $2^m$ by applying the proposed undirected vectors, which increases the efficiency by the square root of the efficiency of the conventional process. Note that reducing the size of the problem reduces the number of connectivity verifications and the number of vector probability calculations.

**4.2 First Connected Vector and Last Disconnected Vector**

In each of the conventional methods (e.g., BAT, BFS, DFS, and UGFM) [44, 50] the search for the binary-state network reliability begins with the vector zero (0, 0, 0, …, 0), whose coordinate values are all zeros, and end with the vector one (1, 1, …, 1), whose coordinate values are all ones. In BFS, DFS, and UGFM [44, 50], we have

$$R = \Pr(\{X \mid, \text{where vector } X \text{ is connected, and } (0, 0, …, 0) \leq X \leq (1, 1, …, 1) \}), \quad (1)$$

and in BAT, we have

$$R = \Pr(\{X \mid, \text{where vector } X \text{ is connected and } (0, 0, …, 0) \underline{<<} X \underline{<<} (1, 1, …, 1) \}). \quad (2)$$

If we can start from a vector larger than vector zero and end at another vector that is less than vector one, then the original vector space between vector zero and vector one can be narrowed to reduce the run time. Therefore, we developed the concepts of the first connected vector ($X_{FC}$) and the last disconnected vector ($X_{LD}$).

$$\begin{aligned} R &= \Pr(\{X \mid \text{vector } X \text{ is connected and } X_{FC} \underline{<<} X << (1, 1, …., 1) \} \\ &= \Pr(\{X \mid \text{vector } X \text{ is connected and } X_{FC} \underline{<<} X << X_{LD} \} \\ &\quad + \Pr(\{X \mid \text{for all vector } X \text{ with } X_{LD} << X \}), \end{aligned} \quad (3)$$

and

$$\begin{aligned} R &= 1 - \Pr(\{X \mid \text{vector } X \text{ is disconnected and } (0, 0, …, 0) << X \underline{<<} X_{LD} \} \\ &= 1 - [\Pr(\{X \mid \text{for all } X \text{ with } X << X_{FC} \}) \\ &\quad + \Pr(\{X \mid \text{vector } X \text{ is disconnected and } X_{FC} << X \underline{<<} X_{LD} \}]. \end{aligned} \quad (4)$$

We can directly calculate $\Pr(\{X \mid \text{for each } X \text{ where } X << X_{FC} \})$ and/or $\Pr(\{X \mid \text{for each vector } X$

where $X_{LD} \ll X$ }) without identifying each vector in {$X$ | for each $X$ where $X \ll X_{FC}$ } and/or Pr({$X$ | for each vector $X$ where $X_{LD} \ll X$ }), as discussed in Section 5.3.

$$\{X \mid \text{vector } X \text{ is disconnected and } X_{FC} \ll X \ll X_{LD} \}$$
$$\subseteq \{X \mid \text{vector } X \text{ is connected and } (0, 0, \ldots, 0) \leq X \leq (1, 1, \ldots, 1)\}$$
$$= \{X \mid \text{vector } X \text{ is connected and } (0, 0, \ldots, 0) \underline{\ll} X \underline{\ll} (1, 1, \ldots, 1)\}. \quad (5)$$

Consequently, the total number of desired vectors is reduced by using the concepts of $X_{FC}$ and $X_{LD}$.

### 4.2.1 First Connected Vector

The basic idea of the $X_{FC}$ is that any vector $X$, where $X \ll X_{FC}$, is disconnected. Based on the fundamental properties of the BAT, $X(1:k) \leq X_{FC}(1:k)$ for some $k < m$, if $X \ll X_{FC}$. If we change $X_{FC}(i)$ from 1 to 0 for any $i$ where $X_{FC}(i) = 1$, it will cause the new $X_{FC}$ to be disconnected. Therefore, the $X_{FC}$ is obtained by finding the forward minimum cuts with the smallest coordinates and changing the state value of the last arc in the selected forward minimum cut to one and the remaining coordinates to zero. For example, (0, 0) is the forward minimum cut with the smallest number of coordinates in Fig. 1 because the other minimum cuts (e.g., $c = \{a_6, a_7\}$) and its maximum arc label is 7, all have coordinates larger than 2. Therefore, let $X_{FC}(1: 2) = (0, 1)$ because (0, 0) has two coordinates.

The above process is repeated, and the $X_{FC}$ can be selected using the following code:

**Algorithm FIND_$X_{FC}$**

**STEP F0.** Let the $X_{FC}$ be vector zero, and $i = 1$.

**STEP F1.** Let the $X_{FC}(1:k_i)$ be a forward minimum cut with the smallest integer $k_i$. If no such $k_i$ exists, go to STEP F3.

**STEP F2.** Let $X_{FC}(k_i) = 1$, $i = i + 1$, and go to STEP F1.

**STEP F3.** $X_{FC}$ is the first connected vector.

It is trivial that the $X_{FC}$ obtained from the above code is a connected vector because there is no cut in $G(X_{FC})$ from STEP F1. If the selected $X_{FC}$ is not the first connected vector, then there is a connected vector $X$, such that $X \ll X_{FC}$. From the important property of the BAT, $X(1:h) \leq X_{FC}(1:h)$ for some positive integer $h$ because $X \ll X_{FC}$.

If $X(j) = 0$ for $j = 1, 2, \ldots, k_1$, then $\{a_1, a_2, \ldots, a_{k_1}\}$ is a minimum cut in $G$, which contradicts the fact that $X$ is a connected vector. If $X(k) = 1 > X_{FC}(k) = 0$ for $k < k_1$, then $X_{FC} \ll X$, which contradicts our assumption that $X \ll X_{FC}$. Therefore, $X(j) = 0$ for $j = 1, 2, \ldots, k_1 - 1$, and $X(k_1) = 1$; $X(1:k_1) = X_{FC}(1:k_1)$.

In the same way, if $X(j) = 0$ for $j = k_1 + 1, k_1 + 2, \ldots, k_2$, then there is a cut in $G(X)$ from STEP F1. If $X(k) = 1 > X_{FC}(k) = 0$ for some $k$, where $k_1 < k < k_2$, then $X_{FC} \ll X$. The two cases described above are impossible, so $X(1:k_2) = X_{FC}(1:k_2)$, and $X = X_{FC}$.

For example, in Fig. 1, where $a_1 = e_{1,2}$, $a_2 = e_{1,3}$, $a_3 = e_{3,5}$, $a_4 = e_{4,5}$, $a_5 = e_{3,4}$, $a_6 = e_{2,4}$, and $a_7 = e_{2,3}$, the $X_{FC} = (0, 1, 0, 0, 0, 1, 0)$, and the procedure for the proposed **Algorithm Find_$X_{FC}$** is listed in Table 4.

**Table 4.** Results and the process for obtaining the $X_{FC}$.

| $i$ | $k_i$ | $c_i$ | old $X_{FC}(1:k)$ | New $X_{FC}(1:k)$ |
|---|---|---|---|---|
| 1 | 2 | $\{a_1, a_2\}$ | (0, 0) | (0, 1) |
| 2 | 6 | $\{a_4, a_5, a_6\}$ | (0, 1, 0, 0, 0, 0) | (0, 1, 0, 0, 0, 1) |
| 3 | | | (0, 1, 0, 0, 0, 1, 0) | (0, 1, 0, 0, 0, 1, 0) |

The decimal equivalent of the binary vector (0, 1, 0, 0, 0, 1, 0) is $1 \times 2^5 + 1 \times 2^1 = 34$. In other words, identifying the $X_{FC}$ allows us to eliminate 34 vectors from the conventional BAT search process because we begin at the $X_{FC}$ rather than at vector zero. Note that each vector $X$ is disconnected when $X \ll X_{FC}$ in the BAT.

The graph shown in Fig. 5 is identical to the graph in Fig. 1 but has different arc labels: $a_1 = e_{1,2}$, $a_2 = e_{1,3}$, $a_3 = e_{3,5}$, $a_4 = e_{4,5}$, $a_5 = e_{3,4}$, $a_6 = e_{2,4}$, and $a_7 = e_{2,3}$. The $X_{FC} = (0, 1, 0, 1, 0, 1, 0)$, as explained in Table 5 and depicted in Fig 6.

**Table 5.** Another results and the process for obtaining the $X_{FC}$.

| $i$ | $k_i$ | $c_i$ | old $X_{FC}(1:k)$ | New $X_{FC}(1:k)$ | Remark |
|---|---|---|---|---|---|
| 1 | 2 | $\{a_1, a_2\}$ | (0, 0) | (0, 1) | See Fig. 6(1) |
| 2 | 4 | $\{a_3, a_4\}$ | (0, 1, 0, 0) | (0, 1, 0, 1) | See Fig. 6(2) |
| 3 | 6 | $\{a_3, a_5, a_6\}$ | (0, 1, 0, 0, 0, 0) | (0, 1, 0, 1, 0, 1) | See Fig. 6(3) |
| 4 | 7 | $\{a_7\}$ | (0, 1, 0, 1, 0, 1, 0) | (0, 1, 0, 1, 0, 1, 1) | |

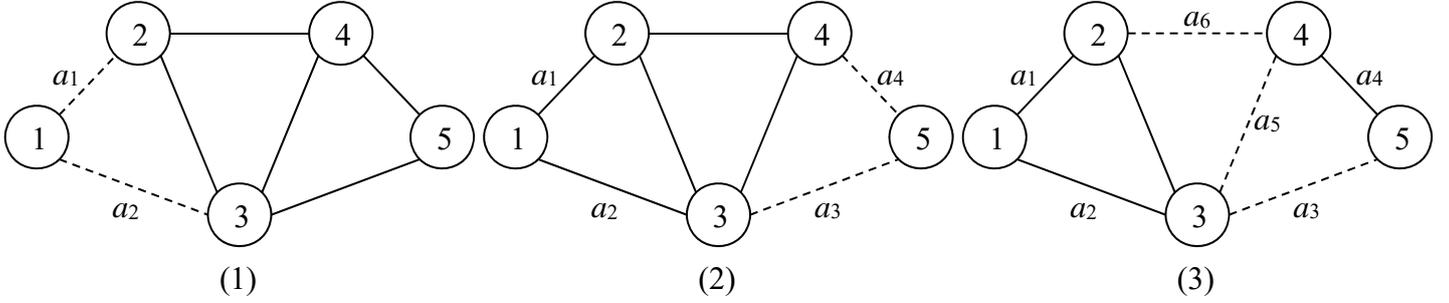

**Figure 6.** Arc labels and finding the $X_{FC}$ for the graph in Fig. 1.

The decimal equivalent of the binary vector $X_{FC} = (0, 1, 0, 1, 0, 1, 0)$ is $1 \times 2^5 + 1 \times 2^3 + 1 \times 2^1$ = 105. Therefore, 105 vectors are eliminated from the conventional BAT search process based on $X_{FC} = (0, 1, 0, 1, 0, 1, 0)$, as shown in Fig. 6.

The two examples above illustrate that the larger *the decimal number of $X_{FC}$*, the more vectors are eliminated from the BAT search process. Moreover, the arc labels affect *the $X_{FC}$*.

**4.2.2 Last Disconnected Vector**

It is easier to identify the $X_{LD}$ than the $X_{FC}$. The $X_{FC}$ is obtained from the first coordinate to the last coordinate, and from multiple minimum cuts found in the sequence. The $X_{LD}$ is identified from one minimum cut and the final coordinate. The code to obtain the $X_{LD}$ is listed below.

**Algorithm FIND_$X_{LD}$**

**STEP L0.** Let $X$ be vector one (i.e., $X(i) = 1$ for all $i = 1, 2, …, m$).

**STEP L1.** Let $k$ be the largest integer such that there is one minimum cut in $G(X(k:m))$.

**STEP L2.** Let $X(i) = 0$ for all $a_i$ in the minimum cut selected in STEP L1.

**STEP L3.** $X = X_{LD}$ is the last disconnected vector.

Because there is a minimum cut in $G(X_{LD})$ from STEP L1, $X_{LD}$ is a disconnected vector. Assume that $X_{LD} << X$, and $X$ is also a disconnected vector. If $X(i) = 0$ for at least one $i \in \{1, 2, \ldots, k–1\}$, then $X << X_{LD}$, which is impossible. If $X(i) = 1$ for at least one $i \in \{k, k+1, \ldots, m\}$, then $X$ is not a disconnected vector because $k$ is the largest integer such that there is a minimum cut in $G(X_{LD}(k:m))$. Using the code listed above, we identify the $X_{LD}$.

For the graph shown in Fig. 1, the $X_{LD} = (1, 1, 1, 1, 1, 0, 0)$, and the procedure for the proposed **Algorithm Find_$X_{LD}$** is listed in Table 6.

Table 6. Results and the process for obtaining the $X_{LD}$ for Fig. 1.

| STEP | k | c | Remark |
|---|---|---|---|
| L0 | | | $X_{LD} = (1, 1, 1, 1, 1, 1, 1)$ |
| L1 | 6 | $\{a_6, a_7\}$ | |
| L2 | | | $X_{LD}(6:7) = (0, 0)$ |
| L2 | | | $X_{LD} = (1, 1, 1, 1, 1, 0, 0)$ |

The decimal equivalent of the difference between binary numbers 1111111 and 1111100 is 4, so identifying the $X_{LD}$ with the proposed algorithm eliminates four vectors because vector $X$ is connected for each $X$ where $X_{LD} << X$.

On the other hand, using the graph in Fig. 5 and the proposed **Algorithm Find_$X_{LD}$**, the $X_{LD} = (1, 1, 0, 1, 0, 0, 1)$, as explained in Table 7.

Table 7. Results and the process for obtaining the $X_{LD}$ for Fig. 5.

| STEP | k | c | Remark |
|---|---|---|---|
| L0 | | | $X_{LD} = (1, 1, 1, 1, 1, 1, 1)$ |
| L1 | 3 | $\{a_3, a_5, a_6\}$ | See Fig. 6(3) |
| L2 | | | $X_{LD}(i) = 0$ for $i = 3, 5, 6$ |
| L2 | | | $X_{LD} = (1, 1, 0, 1, 0, 0, 1)$ |

The decimal equivalent of the difference between binary numbers 1111111 and 1101001 is 23, so identifying the $X_{LD}$ with the proposed algorithm eliminates 23 vectors, representing a more substantial reduction in run time than obtaining the $X_{LD}$ from Fig. 1.

The two examples above illustrate that the smaller the decimal equivalent of $X_{LD}$, the more vectors are eliminated from the BAT search process. Moreover, the arc labels affect the value of the

$X_{LD}$.

## 4.3 Identify the Minimum Cuts First and Label the Arcs Second

The examples presented in Sections 4.2.1 and 4.2.2 show that in Fig. 1, $X_{FC} = X_{34} = (0, 1, 0, 0, 0, 1, 0)$, and $X_{LD} = X_{124} = (1, 1, 1, 1, 1, 0, 0)$, where $a_1 = e_{1,2}$, $a_2 = e_{1,3}$, $a_3 = e_{3,5}$, $a_4 = e_{4,5}$, $a_5 = e_{3,4}$, $a_6 = e_{2,4}$, and $a_7 = e_{2,3}$. In Fig. 5, $X_{FC} = X_{43} = (0, 1, 0, 1, 0, 1, 1)$, and $X_{LD} = X_{105} = (1, 1, 0, 1, 0, 0, 1)$, where $a_1 = e_{1,2}$, $a_2 = e_{1,3}$, $a_3 = e_{3,5}$, $a_4 = e_{4,5}$, $a_5 = e_{3,4}$, $a_6 = e_{2,4}$, and $a_7 = e_{2,3}$. The arc labels in Fig. 1 eliminate (128 – 124) + 34 = 38 vectors (29.68%), and the arc labels in Fig. 5 eliminate (128 – 105) + 43 = 66 vectors (51.56%). Therefore, the arrangement of arc labels is very important in reducing redundant vectors.

Although we have shown that the arc labels can affect the number of eliminated vectors, identifying the arc label arrangement that is optimal for $X_{FC}$ and $X_{LD}$ is difficult. Therefore, we focus on finding the arc labels that maximize the decimal equivalent of $X_{FC}$ to eliminate the maximum number of vectors by first identifying the minimum cuts and then trying to include more minimum cuts in $X_{FC}(1:k)$ for all $k$. The details of this concept are provided in the following procedure:

**Algorithm LABEL_ARCS**

**STEP A0.** Find a minimum cut $c$ such that at least one arc is unlabeled, and then label the unlabeled arcs in $c$.

**STEP A1.** If all arcs are labeled, then halt. Otherwise, go to STEP A1.

For example, in Fig. 6, there are two minimum cuts: $\{e_{1,2}, e_{1,3}\}$ (see Fig. 6(1)) and $\{e_{3,5}, e_{4,5}\}$ (see Fig. 6(2)), and the arcs in these two cuts must be labeled first. We can label $a_1 = e_{1,2}$, $a_2 = e_{1,3}$, $a_3 = e_{3,5}$, and $a_4 = e_{3,4}$. However, we have another minimum cut that has at least one unlabeled arc: $\{a_4 = e_{4,5}, a_5 = e_{3,4}, a_6 = e_{2,4}\}$ (see Fig. 6(3)). In this case, $e_{2,3}$ is labeled as 7 because all the other arcs are labeled. Note that we can also choose the minimum cut $\{a_2 = e_{1,3}, e_{2,4}, e_{2,3}\}$ rather than $\{a_4 = e_{4,5}, a_5$

= $e_{3,4}$, $a_6 = e_{2,4}$} in the above.

## 4.4 Super Vectors

A traditional BAT must wait for a vector search process to complete before determining whether the vector is connected or disconnected. If the connectivity of a sub-vector can be resolved before the search expands to the complete vector, then the run time can be decreased by the time it would take to wait for sub-vectors to become complete vectors. The connectivity of such complete vectors is ensured based on the connectivity of the sub-vectors, which reduces the run time further. The special sub-vector described above is called a super vector and is the novel concept discussed in this section.

A super vector $S$ is a sub-vector that meets the following two conditions:

1. The number of coordinates in $S$ is less than or equal to the number of coordinates in the normal vector (i.e., $|S| \leq |X|$ for all vectors $X$). For example, $S = (0, 0)$ is a super vector in Fig. 1.

2. An object represented by its coordinates is equal to that of all vectors. In other words, the $i$th coordinate represents the state of arc $j$ if the $i$th coordinate also represents the state of arc $j$ in each normal vector for $i = 1, 2, \ldots, |S|$. For example, $S(1) = S(a_1) = 0$ and $S(2) = S(a_2) = 0$ if super vector $S = (0, 0)$ in Fig. 1.

A super vector $S$ is a connected super vector if $G(S)$ is connected, and $G(S(1:|S|-1))$ is not connected. Super vector $S$ is a disconnected vector if $G(S)$ is disconnected, and $G(S(1:|S|-1))$ is not disconnected. Any vector $X$ is a connected or disconnected vector if $X(i) = S(i)$, and the super vector $S$ is connected or disconnected for all $i = 1, 2, \ldots, |S|$.

A normal vector is either connected or disconnected, but a super vector can be connected, disconnected, or neither. Moreover, the value of the last coordinate is 1 in the connected super vectors and 0 in the disconnected super vectors. The last ordinate can be removed from a super vector that is neither connected nor disconnected.

For example, in Fig. 1, the super vector $S = (0, 0)$ is disconnected (see Fig. 6(1)). Each vector $X$ where $X(1) = X(2) = 0$, or $X(1:2) = (0, 0) = S$, is also disconnected. Therefore, the vector next to $S = (0, 0)$ that we need to start generating is $(0, 1, 0, 0, 0, 0, 0)$. In the same way, $S = (0, 1, 0, 0, 0, 1)$ is a connected super vector, and any vector $X$ where $X(1:6) = S$ is also a connected vector. The vector next to $S = (0, 1, 0, 0, 0, 1)$ that we need to start to generating is $(0, 1, 0, 0, 1, 0, 0)$. In addition, the super vector $S = (0, 1)$ is neither connected nor disconnected.

After finding disconnected or connected vectors $S$, there is no need to generate or determine the connectivity of vectors $X$ where $X(1: |S|) = S$. Moreover, we can obtain the vector directly by adding one to the binary $X(1:|S|-1)$ and letting $X(|S|: m) = 0$ if $S$ is connected, or letting $X(|S|) = 1$ and $X(|S|+1: m) = 0$ if $S$ is disconnected. Moreover, Eq.(3) and Eq.(4) can be changed to the following:

$$R = \Pr(\{S \mid \text{super vector } S \text{ is connected and } X_{FC} \underline{\ll} S \ll X_{LD} \})$$
$$+ \Pr(\{X \mid \text{for all vector } X \text{ with } X_{LD} \ll X \}), \qquad (6)$$

and

$$R = 1 - [\Pr(\{X \mid \text{for all } X \text{ with } X \ll X_{FC} \})$$
$$+ \Pr(\{S \mid \text{super vector } S \text{ is disconnected and } X_{FC} \ll S \underline{\ll} X_{LD} \}]. \qquad (7)$$

Because the number of vectors is more than the number of the super vectors, we have

$$\{S \mid \text{super vector } S \text{ is connected and } X_{FC} \underline{\ll} S \ll X_{LD} \}$$
$$\subseteq \{X \mid \text{vector } X \text{ is connected and } X_{FC} \underline{\ll} X \ll X_{LD} \}$$
$$\subseteq \{X \mid \text{vector } X \text{ is connected and } (0, 0, \ldots, 0) \leq X \leq (1, 1, \ldots, 1)\}$$
$$= \{X \mid \text{vector } X \text{ is connected and } (0, 0, \ldots, 0) \underline{\ll} X \underline{\ll} (1, 1, \ldots, 1)\}, \qquad (8)$$

In other words, an algorithm that uses the proposed super vectors is more efficient than an algorithm that does not use them.

## 5. CALCULATING THE RELIABILITY OF NOVEL VECTORS

There are three types of vectors in the proposed bounded BAT: traditional vectors, $X_{FC}$ and $X_{LD}$, and connected or disconnected super vectors. The corresponding (occurrent) probability calculations are also distinctive and are discussed in this section.

### 5.1 Probability of Vectors

Among these three (occurrent) probability calculations, the vector probability is the easiest one to calculate.

$$\Pr(X) = \Pr(\{\text{the state of } a_1 \text{ is } x_1\}) \times \Pr(\{\text{the state of } a_2 \text{ is } x_2\}) \times \ldots$$
$$\times \Pr(\{\text{the state of } a_m \text{ is } x_m\})$$
$$= \prod_{i=1}^{m} \Pr(\{\text{the state of } a_i \text{ is } x_i\}). \tag{9}$$

If all arcs in the binary-state network are homogeneous (i.e., the reliabilities are identical and equal to $r$), then $\Pr(X)$ can be simplified as follows:

$$\Pr(X) = \prod_{i=1}^{m} \Pr(\{\text{the state of } a_i \text{ is } x_i\})$$
$$= r^{\sum_{i=1}^{m} x_i} (1-r)^{m - \sum_{i=1}^{m} x_i}. \tag{10}$$

where $\sum_{i=1}^{m} x_i$ and $(m - \sum_{i=1}^{m} x_i)$ are the numbers of functioning and failed arcs, respectively.

For example, let $X = (0, 1, 0, 1, 0, 1, 1)$ be a vector of a homogeneous binary-state network and $r = 0.8$. We have $\Pr(X) = 0.8^4 \times 0.2^3 = 0.0032768$.

## 5.2 Probability of Connected or Disconnected Super Vectors

The super vector is a special subvector, and the number of its coordinates can be less than $m$. Let $S = (g_1, g_2, ..., g_\gamma)$ be a super vector and $\gamma \leq m$. The coordinates that are not included in $G$ (i.e., coordinates $\gamma+1, \gamma+2, ..., m$) can be any state, so their probabilities are one. Therefore, the calculated probability of $\Pr(S)$ is similar to that of a full vector, except that there is no need to multiply the coordinates that are not included in $S$. The calculation of $\Pr(S)$ can be derived from the following equation:

$$\Pr(S) = \Pr(\{\text{the state of } a_1 \text{ is } g_1\}) \times \Pr(\{\text{the state of } a_2 \text{ is } g_2\}) \times ...$$

$$\times \Pr(\{\text{the state of } a_\gamma \text{ is } g_\gamma\})$$

$$= \prod_{i=1}^{\gamma} \Pr(\{\text{the state of } a_i \text{ is } g_i\}). \tag{11}$$

In the same way as $\Pr(X)$ in a homogeneous binary-state network where the reliability of each arc is $r$, $\Pr(S)$ can be calculated as follows:

$$\Pr(S) = r^{\sum_{i=1}^{\gamma} x_i} (1-r)^{m - \sum_{i=1}^{\gamma} x_i}. \tag{12}$$

For example, assume that $S = (0, 1, 0, 1, 1)$ is a super vector of a homogeneous binary-state network of which $r = 0.8$. Based on Eq.(12), $\Pr(S) = 0.8^3 \times 0.2^2 = 0.02048$.

## 5.3 Probabilities of the Vectors before $X_{FC}$ and after $X_{LD}$

The proposed $X_{FC}$ and $X_{LD}$ are special vectors, and their probability calculations $\Pr(X_{FC})$ and $\Pr(X_{LD})$ are obtained with Eq.(6), as listed in Section 5.1. However, calculating the probabilities of the vectors generated before $X_{FC}$ and after $X_{LD}$ in the BAT is difficult.

Any vector $X$ is disconnected if $X \ll X_{FC}$ or connected if $X_{LD} \ll X$. Therefore, we consider the probability calculations of $\Pr(X_{FC})$ and/or $\Pr(X_{LD})$ and the vectors that were skipped based on $X_{FC}$ and/or $X_{LD}$ (i.e., $\Pr(\{X \mid \text{for all } X \text{ where } X_{LD} \ll X\})$ in Eq.(6) and $\Pr(\{X \mid \text{for all } X \text{ where } X \ll X_{FC}\})$ in Eq.(7)).

Without loss of generality, let $f_i$ be the $i$th 1 appearing in $X_{FC}$, and $i = 1, 2, \ldots, \phi$. For example, $f_1, f_2, f_3, f_4 = 2, 4, 6, 7$ in $X_{FC} = (0, 1, 0, 1, 0, 1, 1)$, and $\phi = 4$. Let $S_{FC,i}$ be a super vector with the following properties:

1. $S_{FC,i}(k) = X_{FC}(k)$ for $k = 1, 2, \ldots, f_i - 1$
2. $S_{FC,i}(i) = 0$
3. $|S_{FC,i}| = i$.

Because

$$\{X \mid \text{for all } X \text{ with } X \ll X_{FC}\} = \{X \mid \text{for all } X \text{ with } X \ll S_{FC,1}\} +$$
$$\{X \mid \text{for all } X \text{ with } S_{FC,1} \underline{\ll} X \ll S_{FC,2}\} + \ldots +$$
$$\{X \mid \text{for all } X \text{ with } S_{FC,\phi-1} \ll X \ll S_{FC,\phi}\}, \quad (13)$$

and any two terms on the right-hand side are disjointed, we have

$$\Pr(X_{FC}) = \Pr(S_{FC,1}) + \Pr(S_{FC,2}) + \ldots + \Pr(S_{FC,\phi}). \quad (14)$$

For example, $S_{FC,1} = (0, 0)$, $S_{FC,2} = (0, 1, 0, 0)$, $S_{FC,3} = (0, 1, 0, 1, 0, 0)$ and $S_{FC,3} = (0, 1, 0, 1, 0, 1, 0)$ if $X_{FC} = (0, 1, 0, 1, 0, 1, 1)$. Hence, if the reliability of each arc is 0.8, then

$$\Pr(\{X \mid \text{for all } X \text{ with } X \ll X_{FC}\}) = \Pr(S_{FC,1}) + \Pr(S_{FC,2}) + \Pr(S_{FC,3}) + \Pr(S_{FC,4})$$
$$= 0.2^2 + 0.2^3 \times 0.8 + 0.2^4 \times 0.8^2 + 0.2^4 \times 0.8^3$$
$$= 0.048243. \quad (15)$$

In the same way, let $f_i$ be the $i$th 1 appearing in $X_{LD}$, and $i = 1, 2, \ldots, \phi$. $S_{LD,i}$ is a super vector with the following properties:

1. $S_{LD,i}(k) = X_{LD}(k)$ for $k = 1, 2, \ldots, f_i - 1$
2. $S_{LD,i}(i) = 1$
3. $|S_{LD,i}| = i$.

We have

$$\{X \mid \text{for all } X \text{ with } X_{LD} \ll X\} = \{X \mid \text{for all } X \text{ with } S_{LD,1} \ll X \ll S_{LD,2}\} +$$
$$\{X \mid \text{for all } X \text{ with } S_{LD,2} \underline{\ll} X \ll S_{LD,3}\} + \ldots +$$
$$\{X \mid \text{for all } X \text{ with } S_{LD,\phi} \ll X\}. \quad (16)$$

In addition, any two terms on the right-hand side are disjointed. Therefore,

$$\Pr(\{X \mid \text{for all } X \text{ with } X \ll X_{FC}\}) = \Pr(S_{FC,1}) + \Pr(S_{FC,2}) + \ldots + \Pr(S_{FC,\phi}) \quad (17)$$

For example, let $X_{LD} = (1, 1, 0, 1, 0, 0)$, we have $S_{LD,1} = (1, 1, 1)$, $S_{LD,2} = (1, 1, 0, 1, 1)$, $S_{LD,3} = (1, 1, 0, 1, 0, 1)$, and

$$\Pr(\{X \mid \text{for all } X \text{ with } X_{LD} \ll X\}) = \Pr(S_{LD,1}) + \Pr(S_{LD,2}) + \Pr(S_{LD,3})$$
$$= 0.8^3 + 0.2 \times 0.8^4 + 0.2^2 \times 0.8^4$$
$$= 0.610304. \quad (18)$$

if $\Pr(a_i) = 0.8$ for $i = 1, 2, \ldots, 7$.

## 6. PROPOSED BOUNDED BAT

The details of how the proposed bounded BAT integrates undirected vectors, $X_{FC}$ and $X_{LD}$, and super vectors is explained in this section using the pseudo-code and experimental analysis.

### 6.1 Pseudo-Code

The pseudo-code of the proposed bounded BAT is given below.

**Algorithm for the Bounded BAT**

**Input:** A binary-state network $G(V, E, \mathbf{D_b})$, where node 1 is the source node, and node $n$ is the sink node.

**Output:** The reliability of $G(V, E, \mathbf{D_b})$.

**STEP 0.** Implement the algorithms proposed in Sections 4.2, 4.3, and 4.4 to find the $X_{FC}$ and $X_{LD}$, label the arcs, and calculate $U = \Pr(\{X \mid \text{for all } X \ll X_{FC}\})$ based on Eq.(14), $X = X_{FC}$, and $i = m$.

**STEP 1.** If $X(a_i) = 1$, let $X(a_i) = 0$, and go to STEP 4.

**STEP 2.** Let $X(a_i) = 1$.

**STEP 3.** Let $i = i - 1$ and return to STEP 1 if $i > 0$.

**STEP 4.** If $X \gg X_{LD}$, halt and $1 - U$ is the reliability of $G(V, E, \mathbf{D_b})$.

**STEP 5.** If $X(1:k)$ is connected for some $k \in \{1, 2, \ldots, m\}$, then let $i = k - 1$, and let $X(k:m)$ be the vector zero; go to STEP 1.

**STEP 6.** Let $i = k - 1$; let $U = U + \Pr(X(1:k))$; let $X(k) = 1$; let $X(k+1:m)$ be the vector zero, and then go to STEP 4, where $k \in \{1, 2, \ldots, m\}$ such that $X(1:k)$ is disconnected.

STEP 0 follows the algorithms proposed in Section 4 to label arcs, find $X_{FC}$ and $X_{LD}$, and calculate the total probability that $X \ll X_{FC}$. STEPs 1 to 4 are a loop like a traditional BAT but with undirected vectors and a new stopping criterion to narrow the vector space: the $X_{LD}$, as shown in STEP 4. STEP 4 also outputs the final reliability if the stopping criterion is met.

STEPs 5 and 6 implement the novel super vectors based on the last paragraph of Section 4.5 to update the connected or disconnected super vectors. In addition, the occurrent probability of the disconnected super vector found in STEP 6 is calculated and added to $U$.

**6.2 Example**

Calculating network reliability is an NP-hard problem [20, 21] and a #P-Hard problem [20, 21]. As such, it is not suitable for demonstrating an algorithm that uses a large example. Therefore, to allow readers to understand the proposed bounded BAT quickly, an intermediate example (Fig. 5) is utilized to illustrate the step-by-step procedure of the proposed algorithm.

**STEP 0.** Based on Section 4., the arcs are labeled: $a_1 = e_{1,2}$, $a_2 = e_{1,3}$, $a_3 = e_{3,5}$, $a_4 = e_{4,5}$, $a_5 = e_{3,4}$, $a_6 = e_{2,4}$, and $a_7 = e_{2,3}$. We then obtain $X = X_{FC} = (0, 1, 0, 1, 0, 1, 1)$; $X_{LD} = (1, 1, 0, 1, 0, 0, 1)$; $U = \Pr(\{X \mid \text{for all } X \ll X_{FC}\}) = 0.0482432$, and $i = 6$.

**STEP 1.** Because $X(a_6) = 1$, let $X(a_6) = 0$.

**STEP 2.** Let $i = i - 1 = 5$ and return to STEP 1 because $i = 5 > 0$.

**STEP 1.** Because $X(a_5) = 1$, let $X(a_5) = 0$.

**STEP 2.** Let $i = i - 1 = 4$ and return to STEP 1 because $i = 4 > 0$.

**STEP 1.** Because $X(a_4) = 0$, go to STEP 3.

**STEP 3.** Let $X(a_4) = 1$.

**STEP 4.** Because $X = (0, 1, 0, 1, 1, 0, 0) \ll X_{LD} = (1, 1, 0, 1, 0, 0, 1)$, go to STEP 5.

**STEP 5.** Because $X(1:5) = (0, 1, 0, 1, 1)$ is connected (i.e., $X(1:5)$ is a connected super vector (Fig. 7(1)), let $i = 4$, and $X(5:7) = (0, 0)$; go to STEP 1.

**STEP 1.** Because $X(a_4) = 1$, let $X(a_4) = 0$.

**STEP 2.** Let $i = i - 1 = 3$ and return to STEP 1 because $i = 3 > 0$.

**STEP 1.** Because $X(a_3) = 1$, go to STEP 3.

**STEP 3.** Let $X(a_3) = 1$.

**STEP 4.** Because $X = (0, 1, 1, 0, 0, 0, 0) \ll X_{LD} = (1, 1, 0, 1, 0, 0, 1)$, go to STEP 5.

**STEP 5.** Because $X(1:3) = (0, 1, 1)$ is a connected super vector (Fig. 7(2)), let $i = 2$, and $X(3:7) = (0, 0, 0, 0, 0)$; go to STEP 1.

**STEP 1.** Because $X(a_2) = 1$, let $X(a_2) = 0$.

**STEP 2.** Let $i = i - 1 = 1$ and return to STEP 1 because $i = 1 > 0$.

**STEP 1.** Because $X(a_1) = 1$, go to STEP 3.

**STEP 3.** Let $X(a_1) = 1$.

**STEP 4.** Because $X = (1, 0, 0, 0, 0, 0, 0) \ll X_{LD} = (1, 1, 0, 1, 0, 0, 1)$, go to STEP 5

**STEP 5.** Because $X(1:4) = (1, 0, 0, 0)$ is a disconnected super vector (Fig. 7(3)), go to STEP 6.

**STEP 6.** Let $i = 3$; let $\Pr(X(1:4)) = 0.0064$; let $U = U + \Pr(X(1:4)) = 0.0546432$; let $X(3:7) = (1, 0, 0, 0, 0)$, and go to STEP 4.

**STEP 4.** Because $X = (1, 0, 1, 0, 0, 0, 0) << X_{LD} = (1, 1, 0, 1, 0, 0, 1)$, go to STEP 5.

**STEP 5.** Because $X$ is a disconnected vector (Fig. 7(4)), go to STEP 6.

**STEP 6.** Let $i = 7$, $\Pr(X) = 0.0002048$, $U = U + \Pr(X) = 0.054848$, and go to STEP 4.

$$\vdots$$

**STEP 4.** Because $X = (1, 1, 0, 1, 0, 0, 0) << X_{LD} = (1, 1, 0, 1, 0, 0, 1)$, go to STEP 5.

**STEP 5.** Because $X(1:6) = (1, 1, 0, 1, 0, 0)$ is a disconnected vector (Fig. 7(5)), go to STEP 6.

**STEP 6.** We have $i = 5$; $\Pr(X(1:6)) = 0.004096$; $U = U + \Pr(X) = 0.0921216$, and $X(6:7) = (1, 0)$; go to STEP 4.

**STEP 4.** Because $X = (1, 1, 0, 1, 0, 1, 0)$ (Fig. 7(6)) $>> X_{LD} = (1, 1, 0, 1, 0, 0, 1)$, halt; and $1 - U = 0.9078784$ is the reliability of $G(V, E, \mathbf{D_b})$.

The results obtained using the proposed bounded BAT are provided in Table 8 and Fig. 7 where dashed arcs are failed, bold arcs are functioning, and the states of other arcs are unknown.

**Table 8.** Final results of using the proposed bounded BAT.

| $i$ | $G_i$ | $C$ | $R_i$ | $i_1$ | $i_2$ | $i$ | $G_i$ | $C$ | $R_i$ | $i_1$ | $i_2$ |
|---|---|---|---|---|---|---|---|---|---|---|---|
|  |  | N | 482432 | 1 | 43 | 14 | (1, 0, 1, 0, 0, 1, 1) | Y | 32768 | 84 | 84 |
| 1 | (0, 1, 0, 1, 0, 1, 1) | Y | 32768 | 44 | 44 | 15 | (1, 0, 1, 0, 1, 0, 0) |  | 8192 | 85 | 85 |
| 2 | (0, 1, 0, 1, 1) | Y | 204800 | 45 | 48 | 16 | (1, 0, 1, 0, 1, 0, 1) | Y | 32768 | 86 | 86 |
| 3 | (0, 1, 1) | Y | 1280000 | 49 | 64 | 17 | (1, 0, 1, 0, 1, 1) | Y | 163840 | 87 | 88 |
| 4 | (1, 0, 0, 0) |  | 64000 | 65 | 72 | 18 | (1, 0, 1, 1, 0, 0, 0) |  | 8192 | 89 | 89 |
| 5 | (1, 0, 0, 1, 0, 0, 0) |  | 2048 | 73 | 73 | 19 | (1, 0, 1, 1, 0, 0, 1) | Y | 32768 | 90 | 90 |
| 6 | (1, 0, 0, 1, 0, 0, 1) |  | 8192 | 74 | 74 | 20 | (1, 0, 1, 1, 0, 1, 0) | Y | 32768 | 91 | 91 |
| 7 | (1, 0, 0, 1, 0, 1) | Y | 40960 | 75 | 76 | 21 | (1, 0, 1, 1, 0, 1, 1) | Y | 131072 | 92 | 92 |
| 8 | (1, 0, 0, 1, 1, 0, 0) |  | 8192 | 77 | 77 | 22 | (1, 0, 1, 1, 1, 0, 0) |  | 32768 | 93 | 93 |
| 9 | (1, 0, 0, 1, 1, 0, 1) | Y | 32768 | 78 | 78 | 23 | (1, 0, 1, 1, 1, 0, 1) | Y | 131072 | 94 | 94 |
| 10 | (1, 0, 0, 1, 1, 1) | Y | 163840 | 79 | 80 | 24 | (1, 0, 1, 1, 1, 1) | Y | 655360 | 95 | 96 |
| 11 | (1, 0, 1, 0, 0, 0, 0) |  | 2048 | 81 | 81 | 25 | (1, 1, 0, 0) |  | 256000 | 97 | 104 |
| 12 | (1, 0, 1, 0, 0, 0, 1) | Y | 8192 | 82 | 82 | 26 | (1, 1, 0, 1, 0, 0) |  | 40960 | 105 | 106 |
| 13 | (1, 0, 1, 0, 0, 1, 0) |  | 8192 | 83 | 83 |  |  | Y | 6103040 | 107 | 128 |

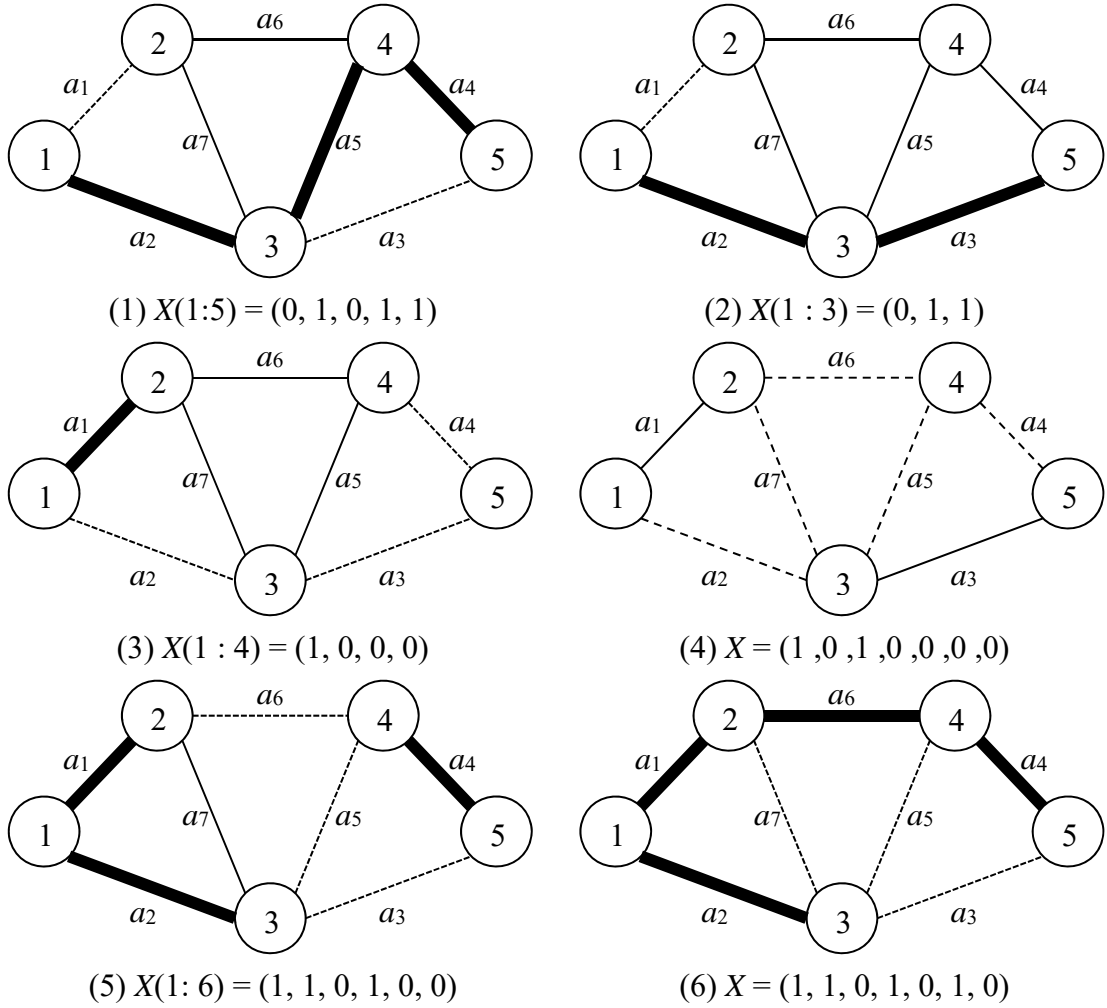

(1) $X(1:5) = (0, 1, 0, 1, 1)$  
(2) $X(1:3) = (0, 1, 1)$  
(3) $X(1:4) = (1, 0, 0, 0)$  
(4) $X = (1, 0, 1, 0, 0, 0, 0)$  
(5) $X(1:6) = (1, 1, 0, 1, 0, 0)$  
(6) $X = (1, 1, 0, 1, 0, 1, 0)$  

**Figure 7.** Example network explaining the proposed new BAT.

From the above, the number of vectors obtained using the proposed bounded BAT is only 26, which is 21.875% of the number obtained using a traditional BAT with the proposed undirected vectors (Table 3) and 2.734% of the number obtained with a conventional BAT. Therefore, each of the proposed novel components (i.e., $X_{FC}$, $X_{LD}$, super vectors, and undirected vectors) are useful for improving the efficiency of the BAT.

## 7. CONCLUSIONS

Network structures and models have been widely adopted, and many are based on the binary-state network. Reliability is the most commonly used tool to evaluate network performance.

In this study, a new bounded BAT is proposed to improve the efficiency of calculating the reliability of a binary-state network. Novel concepts introduced with the proposed bounded BAT include undirected vectors, the first connected vector, the last disconnected vector, and disconnected or connected super vectors. These concepts narrow the vector space to reduce the problem size.

In future works, the proposed BAT will be improved by considering a better and dynamic way to relabel arcs. The proposed BAT will be extended to additional real-life applications with larger problem sizes by incorporating artificial intelligence or Monte Carlo simulations.

**ACKNOWLEDGEMENTS**

This research was supported in part by the Ministry of Science and Technology, R.O.C. under grant MOST 107-2221-E-007-072-MY3 and MOST 104-2221-E-007-061-MY3.

**References**

[1] W.C. Yeh and J.S. Lin, "New parallel swarm algorithm for smart sensor systems redundancy allocation problems in the Internet of Things," *The Journal of Supercomputing*, vol. 74, no. 9, pp. 4358-4384, 2018.

[2] H. Lin and M. Tseng, "Two-level, multistate Markov model for satellite propagation channels," *IEE Proceedings-Microwaves, Antennas and Propagation,* vol. 151, no. 3, pp. 241-248, 2004.

[3] X. Song, X. Jia, and N. Chen, "Sensitivity Analysis of Multi-State Social Network System Based on MDD Method," *IEEE Access,* vol. 7, pp. 167714-167725, 2019.

[4] W.C. Yeh, "A Novel Generalized Artificial Neural Network for Mining Two-Class Datasets," *arXiv preprint arXiv:1910.10461*, 2019.

[5] W.C. Yeh, "A squeezed artificial neural network for the symbolic network reliability functions of binary-state networks," *IEEE transactions on neural networks and learning systems*, vol. 28, pp. 2822-2825, 2016.

[6] D. Kakadia and J. E. Ramirez-Marquez, "Quantitative approaches for optimization of user experience based on network resilience for wireless service provider networks," *Reliability Engineering & System Safety*, vol.193, ID. 106606, 2020.

[7] C. Lin, L. Cui, D.W. Coit and M. Lv, "Performance analysis for a wireless sensor network of star topology with random nodes deployment," *Wireless Personal Communications*, vol. 97, pp. 3993-4013, 2017.

[8] M. Wang, W.C. Yeh, T.C. Chu, X. Zhang, C.L. Huang, and J. Yang, "Solving Multi-Objective Fuzzy Optimization in Wireless Smart Sensor Networks under Uncertainty Using a Hybrid of IFR and SSO Algorithm," *Energies*, https://doi.org/10.3390/en11092385, 2018.

[9] Á. Rodríguez-Sanz, D. Á. Álvarez, F. G. Comendador, R. A. Valdés, J. Pérez-Castán, and M. N. Godoy, "Air Traffic Management based on 4D Trajectories: A Reliability Analysis using Multi-State Systems Theory," *Transportation research procedia,* vol. 33, pp. 355-362, 2018.

[10] P. Wang, R. Billinton, and L. Goel, "Unreliability cost assessment of an electric power system using reliability network equivalent approaches," *IEEE Transactions on power systems,* vol. 17, no. 3, pp. 549-556, 2002.

[11] W.C. Yeh, Y. Jiang, C.L. Huang, N.N. Xiong, C.F. Hu, and Y.H. Yeh, "Improve Energy Consumption and Signal Transmission Quality of Routings in Wireless Sensor Networks," *IEEE Access*, 2020.

[12] W. C. Yeh, "An Improved Method for the Multistate Flow Network Reliability with Unreliable Nodes and the Budget Constraint Based on Path Set", *IEEE Transactions on Systems*, *Man, and Cybernetics: Systems* (renamed from *IEEE Transactions on Systems, Man, and Cybernetics -- Part A: Systems and Humans*), vol. 41, no. 2, pp. 350-355, 2011.

[13] S. Laitrakun and E. J. Coyle, "Reliability-based splitting algorithms for time-constrained distributed detection in random-access WSNs," *IEEE Transactions on Signal Processing,* vol. 62, no. 21, pp. 5536-5551, 2014.

[14] J. E. Ramirez-Marquez, "Assessment of the transition-rates importance of Markovian systems at steady state using the unscented transformation," *Reliability Engineering & System Safety*, vol. 142, pp. 212-220, 2015.

[15] C. M. R. Sanseverino and J. E. Ramirez-Marquez, "Uncertainty propagation and sensitivity analysis in system reliability assessment via unscented transformation," *Reliability Engineering & System Safety*, vol. 132, pp. 176-185, 2014.

[16] W. C. Yeh, "Orthogonal Simplified Swarm Optimization for the Series-Parallel Redundancy Allocation Problem with a Mix of Components," *Knowledge-Based Systems, vol.* 64, pp. 1-12, 2014.

[17] C. L. Huang, "A particle-based simplified swarm optimization algorithm for reliability redundancy allocation problems," *Reliability Engineering & System Safety*, vol. 142, pp. 221-230, 2015.

[18] W. C. Yeh, "A Novel Boundary Swarm Optimization for Reliability Redundancy Allocation Problems", *Reliability Engineering & System Safety*, vol. 192, ID. 106060, 2019.

[19] W. C. Yeh, "A New Exact Solution Algorithm for a Novel Generalized Redundancy Allocation Problem," *Information Sciences*, vol. 408, pp. 182-197, 2017.

[20] D. Shier, *Network Reliability and Algebraic Structures*, Clarendon Press, New York, NY, USA, 1991.

[21] C. J. Colbourn, *The combinatorics of network reliability*, Oxford University Press, 1987.

[22] G. Levitin, *The universal generating function in reliability analysis and optimization*, Springer, 2005.

[23] D.W. Coit and E. Zio, "The evolution of system reliability optimization," *Reliability Engineering & System Safety*, vol. 192, ID. 106259. 2018.

[24] Z. Hao, W. C. Yeh, J. Wang, G. G. Wang and B. Sun, "A quick inclusion-exclusion technique," *Information Sciences*,vol. 486, pp. 20-30, 2019.

[25] S. Chakraborty, N. K. Goyal, S. Mahapatra, and S. Soh, "Minimal Path-Based Reliability Model for Wireless Sensor Networks With Multistate Nodes," *IEEE Transactions on Reliability,* vol. 69, no. 1, pp. 382-400, 2019.

[26] W. C. Yeh, "A Novel Node-based Sequential Implicit Enumeration Method for finding all d-MPs in a Multistate Flow Network," *Information Sciences*, vol. 297, pp. 283-292, 2015.

[27] W. C. Yeh, "A Greedy Branch-and-Bound Inclusion-Exclusion Algorithm for Calculating the Exact Multi-State Network Reliability", *IEEE Transactions on Reliability*, vol.57, no. 1, pp.88-93, 2008.

[28] W. C. Yeh, "An improved sum-of-disjoint-products technique for symbolic multi-state flow network reliability," *IEEE Transactions on Reliability*, vol. 64, no. 4, pp. 1185-1193, 2015.

[29] M. J. Zuo, Z. Tian, and H. Z. Huang, "An efficient method for reliability evaluation of multistate networks given all minimal path vectors", *IIE Transactions*, vol. 39, no. 8, pp. 811-817, 2007.

[30] G. Bai, M. J. Zuo, and Z. Tian, "Search for all *d*-MPs for all *d* levels in multistate two-terminal networks," *Reliability Engineering & System Safety*, vol. 142, pp. 300-309, 2015.

[31] Y. Niu, Z. Gao, and H. Sun, "An improved algorithm for solving all *d*-MPs in multi-state networks," *Journal of Systems Science and Systems Engineering*, vol. 26, no. 6, pp, 711-731, 2017.

[32] G. Levitin, L. Xing, and Y. Dai, "Optimal Spot-Checking for Collusion Tolerance in Computer Grids," *IEEE Transactions on Dependable and Secure Computing*, vol. 16, no. 2, pp. 301-312, 2017.

[33] W. C. Yeh and M. El Khadiri, "A New Universal Generating Function Method for Solving the Single (*d*,τ)-Quick path Problem in Multistate Flow Networks", *IEEE Transactions on Systems, Man, and Cybernetics: Systems* (renamed from *IEEE Transactions on Systems, Man, and Cybernetics -- Part A: Systems and Humans*)*,* vol. 42, no. 6, pp. 1476-1484, 2012.

[34] W. C. Yeh, "A Simple Minimal Path Method for Estimating the Weighted Multi-Commodity Multistate Unreliable Networks Reliability," *Reliability Engineering & System Safety,* vol. 93, no. 1, pp. 125-136, 2008.

[35] W. C. Yeh, L. E. Lin, Y. C. Chou, and Y. C. Chen, "Optimal Routing for Multi-commodity in Multistate Flow Network with Time Constraints", *Quality Technology & Quantitative Management*, vol. 10, no. 2, pp. 161-177, 2013.

[36] Z. Hao, W. C. Yeh, Mi. Zuo, and J. Wang, "Multi-distribution multi-commodity multistate flow network model and its reliability evaluation algorithm", *Reliability Engineering & System Safety*, vol. 193, ID. 106668, 2020.

[37] W. C. Yeh, Y. C. Lin, and Y. Y. Chung, "Performance analysis of cellular automata Monte Carlo Simulation for estimating network reliability," *Expert Systems with Applications*, vol. 37, no. 5, pp. 3537-3544, 2010.

[38] Y. F. Niu and F. M. Shao, "A practical bounding algorithm for computing two-terminal reliability based on decomposition technique," *Computers & Mathematics with Applications*, vol. 61, no. 8, pp. 2241-2246, 2011.

[39] R. E. Bryant, "Graph-based algorithms for boolean function manipulation," *Computers, IEEE Transactions on*, vol. 100, no. 8, pp. 677-691, 1986.

[40] C. Y. Lee, "Representation of switching circuits by binary-decision programs," *The Bell System Technical Journal*, vol. 38, no. 4, pp. 985-999, 1959.

[41] W. C. Yeh, C. Bae, and C. L. Huang, "A new cut-based algorithm for the multi-state flow network reliability problem," *Reliability Engineering & System Safety*, vol. 136, pp.1-7, 2015.

[42] W.C. Yeh, "Search for MC in modified networks," *Computers & Operations Research*, vol. 28, no. 2, pp. 177-184, 2001.

[43] W. C. Yeh, "A simple algorithm to search for all MCs in networks," *European Journal of Operational Research*, vol. 174, no. 3, pp. 1694-1705, 2006.

[44] W. C. Yeh, "Novel Binary-Addition Tree Algorithm (BAT) for Binary-State Network Reliability Problem," *arXiv preprint arXiv:2004.08238*, 2020.

[45] Y. Z. Su and W. C. Yeh, "Binary-Addition Tree Algorithm-Based Resilience Assessment for Binary-State Network Problems," *Electronics*, vol. 9, no. 8, pp. 1207, 2020.

[46] W. C. Yeh, X. Y. Zhang, C. L. Huang, N. N. Xiong, Y. H. Yeh, and C. Jian, "Predicting and Modeling Wildfire Propagation Areas with BAT and Maximum-State PageRank," *IEEE Access*, 2[nd] revision, 2020.

[47] W. C. Yeh, W. B. Zhu, and N. N. Xiong, "Predicting Spread Probability of Learning-Effect Computer Virus," *IEEE Access*, 2[nd] revision, 2020.

[48] E. W. Dijkstra, "A note on two problems in connexion with graphs," *Numerische Mathematik*, vol. 1, pp. 269–271, 1959.

[49] https://en.wikipedia.org/wiki/Stoer-Wagner_algorithm

[50] Z. F. Hao, W. C. Yeh, and M. Forghani-elahabad, "Novel Binary-Addition Tree Algorithm for Reliability Evaluation of Acyclic Multistate Information Networks," *Reliability Engineering & System Safety*, under review, 2020.